\documentclass[aps,prl,reprint,amsmath,amssymb,onecolumn,12pt]{revtex4-2}
\usepackage[letterpaper, left=0.8in, right=0.8in, top=0.9in, bottom=0.9in]{geometry}

\usepackage{cmap} 
\usepackage[utf8]{inputenc}
\usepackage[english]{babel}
\usepackage[T1]{fontenc}
\usepackage{amsmath}
\usepackage{amsfonts}
\usepackage{bm}
\usepackage{xcolor}
\definecolor{blueviolet}{rgb}{0.2, 0.2, 0.6}
\definecolor{webgreen}{rgb}{0,.5,0}
\definecolor{webbrown}{rgb}{.6,0,0}
\usepackage[pdftex,
	bookmarks=false,
	colorlinks=true, 
	urlcolor=webbrown, 
	linkcolor=blueviolet, 
	citecolor=webgreen,
	pdfstartpage=1,
	pdfstartview={FitH},  
	bookmarksopen=false
	]{hyperref}
\usepackage{tikz}
\usepackage{natbib}
\usepackage{mathtools}
\usepackage{graphicx}
\usepackage{csquotes}
\usepackage{braket}
\usepackage{dsfont}
\usepackage{listings}
\allowdisplaybreaks

\usepackage{multirow}
\usepackage{amssymb}

\usepackage{amsthm}

\usepackage{color}
\usepackage{nicefrac}

\DeclareFixedFont{\ttb}{T1}{txtt}{bx}{n}{9} 
\DeclareFixedFont{\ttm}{T1}{txtt}{m}{n}{9}  

\usepackage{color}
\definecolor{deepblue}{rgb}{0,0,0.5}
\definecolor{deepred}{rgb}{0.6,0,0}
\definecolor{deepgreen}{rgb}{0,0.5,0}

\newcommand\pythonstyle{\lstset{
language=Python,
basicstyle=\ttm,
morekeywords={self},              
keywordstyle=\ttb\color{deepblue},
emph={MyClass,__init__},          
emphstyle=\ttb\color{deepred},    
stringstyle=\color{deepgreen},
frame=tb,                         
showstringspaces=false
}}

\newcommand\pythoninline[1]{{\pythonstyle\lstinline!#1!}}

\definecolor{orange}{RGB}{255,127,0}

\def\ket#1{\ensuremath{\mathinner{|{#1}\rangle}}}

\newcommand{\norm}[1]{\left\lVert#1\right\rVert}

\newcommand{\tr}{\text{Tr}}

\newtheorem{theorem}{Theorem}
\newtheorem{corollary}{Corollary}

\usepackage[noend]{algpseudocode}
\usepackage{algorithm,algorithmicx}

\algrenewcommand\alglinenumber[1]{\sf\scriptsize\color{blue}{#1}}
\algrenewcommand\algorithmicrequire{\textbf{Input:}}
\algrenewcommand\algorithmicensure{\textbf{Output:}}

\begin{document}

\title{What if you have only one copy?\\$\&$ Low-depth quantum circuits have no advantage in  decision problems!}

\author{Nengkun Yu}
    \email{nengkun.yu@stonybrook.edu}
	\affiliation{Computer Science Department,
 Stony Brook University}

\date{\today}

\begin{abstract}
The conventional approach to understanding the characteristics of an unknown quantum state involves having numerous identical independent copies of the system in that state. However, we demonstrate that gleaning insights into specific properties is feasible even with a single-state sample. Perhaps surprisingly, the confidence level of our findings increases proportionally with the number of qubits. Our conclusions apply to quantum states with low circuit complexity, including noise-affected ones. Additionally, this extends to learning from a solitary sample of probability distributions. Our results establish a strong lower bound for discriminating quantum states with low complexity. Furthermore, we reveal no quantum advantage in decision problems involving low-depth quantum circuits. Our results can be used to verify NISQ devices.
\end{abstract}

\maketitle

\section{Introduction}
As experiments advance to manipulate larger quantum systems, it becomes crucial to comprehend the streamlined and enhanced efficiency in extracting essential information. This understanding is pivotal for driving progress in theoretical insights and practical applications within quantum information processing, especially in quantum hardware verification and validation. Quantum tomography is crucial in advancing our understanding of quantum phenomena, fostering breakthroughs across diverse scientific and technological frontiers. Quantum state tomography describes a quantum system's state via measurement analysis, which is crucial for verifying computations and channel characterization.
A drawback of quantum state tomography is its impracticality for large-scale systems due to exponential resource requirements \cite{BBMR04,Keyl06,GJK08,HHJ+16,OW16,OW17,FlammiaGrossLiuEtAl2012}.

Focusing on parts of quantum information \cite{10.1145/3188745.3188802}, methods like quantum overlapping tomography efficiently capture reduced density matrices rather than entire quantum states \cite{Cotler_2020}. The classical shadows predict multiple functions of a quantum state using just a logarithmic number of measurements, opening pathways for measuring many-body correlations and entanglement \cite{Huang_2020,Huang_2022,evans2019scalable,10.1145/3406325.3451109,https://doi.org/10.48550/arxiv.2009.04610}. This research breakthrough facilitates efficient measurement of many-body correlations and entanglement.

In \cite{yu2023learning}, we establish the connection between the circuit complexity of quantum states and the sample complexity of learning. Specifically, we demonstrate that states with low circuit complexity can be tomographically reconstructed using a polynomial number of samples and Pauli measurements. Furthermore, \cite{zhao2023learning} addresses the tomography challenge of circuits with bounded gate complexity by employing general measurements. Additionally, \cite{huang2024learning} resolves the tomography of shallow quantum circuits in terms of diamond norm using Pauli measurements. 

\begin{figure}[h!]
\includegraphics[width=8cm]{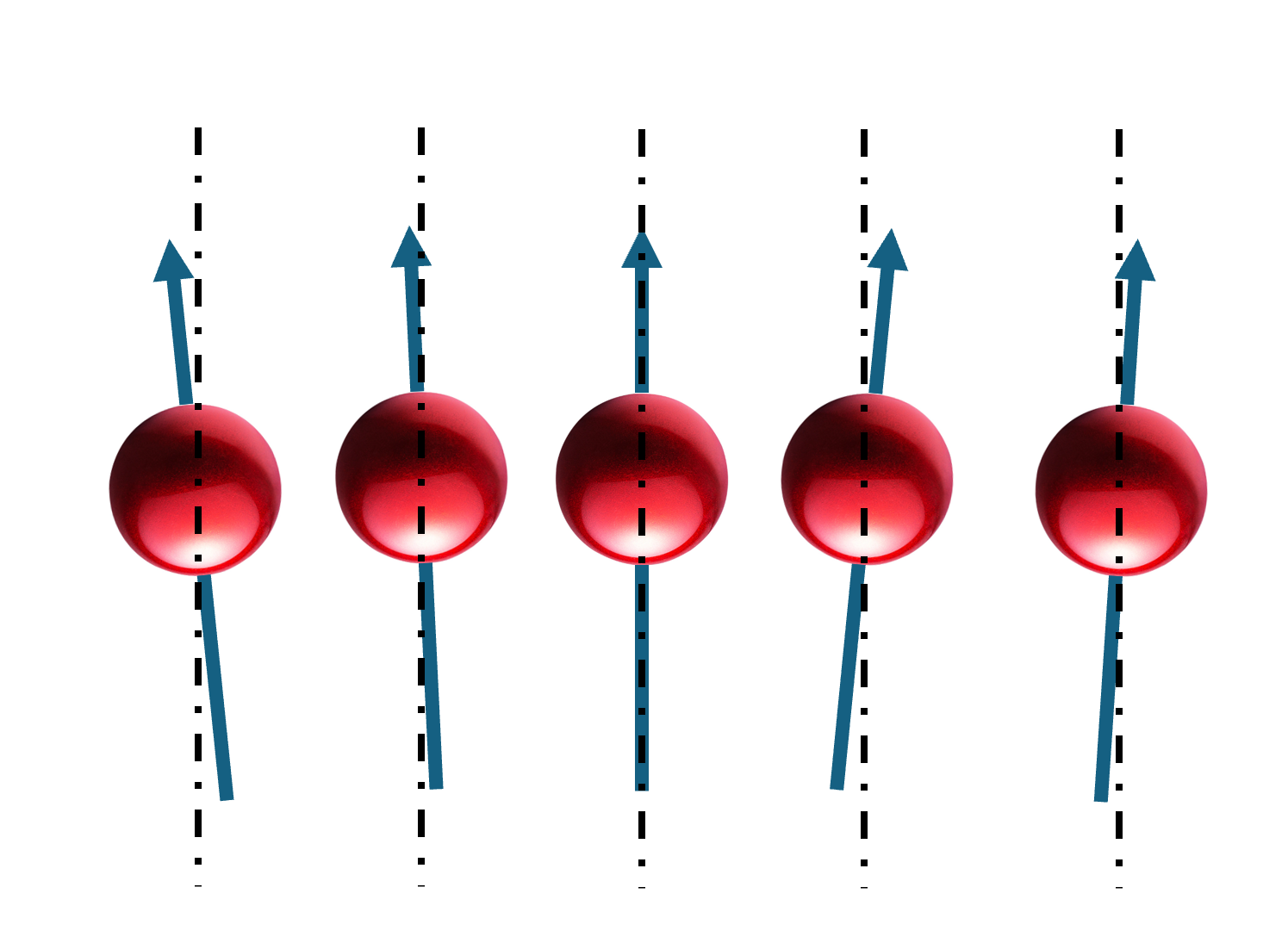}
\centering
\caption{Not IID copies}
\end{figure}

In the works mentioned above, nearly all, if not all, depend on the fundamental requirement of accessing independent identical copies of $\rho$ within a $d$-dimensional Hilbert space. The requirement for identical independent samples stems from utilizing statistical tools such as the Markov inequality or the central limit theorem. These tools are pivotal in quantum tomography, enabling robust statistical inference and reconstruction of quantum states or operations from experimental data. However, this premise tends to be overly idealistic in real-world experimental settings, where quantum noise and coherence are inevitably present, complicating the attainment of perfect replicas.

Is it feasible to learn valuable information without relying on the assumption of identical and independent samples? 

\begin{figure}[h!]
\includegraphics[width=15cm]{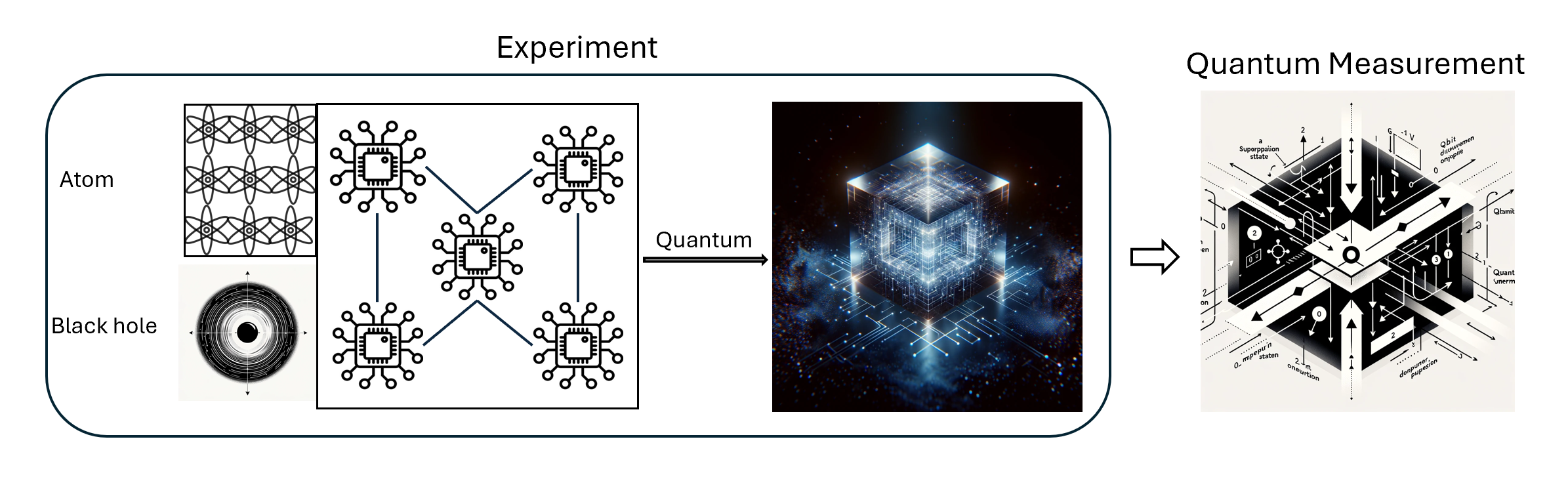}
\centering
\caption{Is learning with a single sample feasible?}
\end{figure}

If we are provided with only a single copy of the quantum state, we don't need to make any assumptions regarding the independence or identity of multiple copies. On the other hand, the setting of a single copy prevents us from getting high confidence in learning by directly utilizing statistical tools. First, is learning some information from a single copy always possible? Intuitively, managing the statistical fluctuation of samples concerning a single reduced-density matrix appears impossible due to its inherent variability. This observation motivates us to consider the average of many reduced density matrices, for example, $\frac{1}{n}\sum Z_i$. Unfortunately, this seems too good to be true within a single copy of a general state:
Consider a class of GHZ states $\ket{\psi_x}=x\ket{0}^{\otimes n}+\sqrt{1-x^2}\ket{1}^{\otimes n}$ with $0\leq x\leq 1$. To learn $\frac{1}{n}\sum Z_i$, we shall measure $Z$ basis on each qubit to get an unbiased estimation. Then, with probability $x^2$, the measurement outcome is $0\cdots 0$; with probability $1-x^2$, the measurement outcome is $1\cdots 1$. One can not learn information of $x$ through measuring one copy of $\ket{\psi_x}$, and therefore can not estimate $\frac{1}{n}\sum Z_i$.
This example suggests that we should consider quantum states that are not highly entangled. 

This is the case when we focus on states that can be generated from low-depth quantum circuits, denoted as low circuit complexity. We will focus on Pauli measurement to measure marginal information With a single copy of the state. For any Pauli matrix $P:=P_1\otimes \cdots \otimes P_n \in \{\sigma_X,\sigma_Y,\sigma_Z\}^{\otimes n}$, and any $s_i\subset\{1,2,\cdots,n\}$, we denote $P_{s_i}=\otimes_{j\in s_{i}}P_{j}$.
We denote the set of observables that can be implemented by measuring $P$ as $\mathcal{O}_P:=\{\sum p_{s_i} P_{s_i}|\sum |p_{s_i}|=1, p_{s_i}\neq 0\}$. For $O=p_{s_i} P_{s_i}\in \mathcal{O}_P$, we call $\max_i |s_i|$ as the length of $O$, and the degree of $O$ as $\max_i |\{s_j|s_i\cap s_j\neq \emptyset\}|$. We also define the norm of $O$, denoted as $\norm{O}$, as $\sum_i p_{s_i}^2$.

\begin{theorem}
One can estimate $\mathrm{tr}(O\rho)$ with good precision and high confidence for a single copy of an unknown quantum state $\rho$ with low circuit complexity if $O\in \mathcal{O}_P$, for some $P\in \{\sigma_X,\sigma_Y,\sigma_Z\}^{\otimes n}$, is of constant length, constant degree and small norm, $\norm{O}=o(1)$.
\end{theorem}
We provide the proof sketch to show that empirical estimation leads to good estimation. We first illustrate this result for pure state $\ket{\psi}=\Pi_{j=1}^k U_j\ket{0}^{\otimes n}$ with constant $k$ and each $U_j$ being a layer of unitary, i.e., the tensor product of two-qubit unitaries apply on disjoint sets of two qubits. 
\begin{align*}
\tr(O\psi)=\sum p_{s_i} \langle \psi|P_{s_i}\ket{\psi}=\sum p_{s_i} \langle 0^{\otimes n}|\Pi_{j=k}^1 U_j^{\dag} P_{s_i}\Pi_{j=1}^k U_j\ket{0^{\otimes n}}
\end{align*}
Consider each term $a_i:=\langle 0^{\otimes n}|\Pi_{j=k}^1 U_j^{\dag} P_{s_i}\Pi_{j=1}^k U_j\ket{0^{\otimes n}}$, we know that $\Pi_{j=k}^1 U_j^{\dag} P_{s_i}\Pi_{j=1}^k U_j$ is an observable only apply non-trivially on a set of qubits, denoted as $t_i$, where the number of qubits depends on the geometrical structure of the circuit and $s_i$, at most $2^k |s_i|$. Focusing on constant $k$ and constant length $O$, we know that each 
$t_i$ is a constant. One can further show that $\max_i |\{t_j|t_i\cap t_j\neq \emptyset\}|$ is a constant by induction. We only need to demonstrate for the first layer, $U_1$: for each $s_i$, there will be at most a constant number of $s_j$ such that $s_i\cap s_j=\emptyset$ and after $U_1$, they become to share some common qubits.

We use $R_i$ to denote the random variable corresponding to the empirical estimation of $A_i$. Clearly, $R_i$ and $R_j$ are independent if $t_i\cap t_j=\emptyset$. We denote random variable $Y:=\sum p_{s_i} R_i$, then $\mathbb E Y=\tr(O\psi)$. We can upper bound the variance of $Y$
\begin{align*}
\mathrm{Var}(Y)=\mathbb{E}(Y-\mathbb{E}Y)^2=\mathbb{E}(\sum p_{s_i} R_i-\sum p_{s_i} a_i)^2=\sum_{i,j} p_{s_i}p_{s_j} \mathbb{E}(R_i-a_i)(R_j-a_j).
\end{align*}
The key observation is that $\mathbb{E}(R_i-a_i)(R_j-a_j)=0$ for $t_i\cap t_j=\emptyset$. This leads to 
\begin{align*}
\mathrm{Var}(Y)=\sum_{t_i\cap t_j\neq\emptyset} p_{s_i}p_{s_j} \mathbb{E}(R_i-a_i)(R_j-a_j)\leq C\sum_{t_i\cap t_j\neq\emptyset} (p_{s_i}^2+p_{s_j}^2)\leq C' (\sum_i p_{s_i}^2)=o(1),
\end{align*}
where $C$ is an upper bound of $|\mathbb{E}(R_i-a_i)(R_j-a_j)|$, $C'/C$ is an upper bound of $\max_i |\{t_j|t_i\cap t_j\neq \emptyset\}|$.

According to Chebyshev inequality, we know that, for any $\epsilon>0$, 
\begin{align*}
\mathrm{Pr}(|Y-\tr(O\psi)|\geq \epsilon)\leq \frac{\mathrm{Var}(Y)}{\epsilon^2}=\frac{o(1)}{\epsilon^2}.
\end{align*}

According to the union bound, one can learn $O_1,\cdots, O_m\in \mathcal{O}_P$ upto $\epsilon$ as long as $\sum \norm{O_i}=o(\epsilon^2)$.

The above conclusion seems highly counter-intuitive:
The number of required samples grows according to the system size in all existing work for quantum learning tasks. Here, if one aims to learn some nontrivial information, one sample could be enough for a system with a more significant number of qubits. At the same time, such a task becomes infeasible in quantum systems with only a few qubits. 

The aforementioned result can be readily extended to low-depth quantum operations, where each layer involves the tensor product $\otimes \mathcal{E}^{i,j}$, with $\mathcal{E}^{i,j}$ representing a two-qubit completely positive trace-preserving map on qubits $i$ and $j$. This extension is achieved by introducing multiple ancilla qubits and employing Stinespring extension on each $\mathcal{E}^{i,j}$ operation, utilizing distinct ancilla qubits.

We have a direct consequence of the probabilistic version. We can define the linear function of distributions similarly to observables, as well as the degree, length and norm of the linear functions.

\begin{corollary}
Let $q$ an unknown distribution on $A_1\times A_2\times\cdots\times A_n$, and it can be generated by applying a constant layer of local Markov chains $\otimes M_{i,j}$ on $(1,0,\cdots,0)$ where any $1\leq i\leq n$ only appears at most once within each layer. 
One can estimate $\tr(Lq)$ with good precision and high confidence for a single sample of $q$ for $L$ being constant length, degree and small norm.
\end{corollary}

Our findings apply to discriminating between quantum states with low circuit complexity. 
\begin{corollary}
For two quantum states $\rho$ and $\sigma$ with constant circuit complexity and $O=\sum \frac{1}{n}Z_i$, $|\mathrm{tr}(O\rho)-\mathrm{tr}(O\sigma)|>\epsilon$ implies $\frac{1}{2}\norm{\rho-\sigma}\geq 1-O(\frac{1}{n\epsilon^2})$.
\end{corollary}
Consider the state discrimination of $\rho$ and $\sigma$ with a uniform prior. Applying our result, one can estimate the value of $O$ for the selected state up to precision $\epsilon/2$ with a failure probability of at most $O(\frac{1}{n\epsilon^2})$. This precision is enough to distinguish $\rho$ and $\sigma$. Therefore, our failure probability of discrimination is at most $O(\frac{1}{n\epsilon^2})$. On the other hand, according to the Helstrom bound, we have $\frac{1-\frac{1}{2}\norm{\rho-\sigma}}{2}\leq O(\frac{1}{n\epsilon^2})$. This is the conclusion.

The above result is also applicable to other observables.

Though one can grasp the properties of low-complexity quantum states from a single sample, the output state quickly spirals into incomprehensibility as the circuit depth deepens, even for depth $4$ circuit ~\cite{https://doi.org/10.48550/arxiv.quant-ph/0205133}. Watrous showed that estimating a single qubit outcome of a general circuit
is complete for quantum statistical zero-knowledge
in \cite{10.5555/645413.652197}. One may wonder whether estimating a single qubit outcome of a low-depth circuit is still hard.

The following result shows that constant depth quantum computation provides no advantage in decision problems. 
For any Boolean function $f:\{0,1\}^n\mapsto \{0,1\}$, we say that a quantum circuit $U$ can compute this function if $\mathrm{tr}(\psi_1 \ket{0}\langle 0|)>2/3$ for $f(i_1,\cdots,i_n)=0$ and $\mathrm{tr}(\psi_1 \ket{0}\langle 0|)<1/3$ for $f(i_1,\cdots,i_n)=1$ are valid for all $i_1,\cdots,i_n$, where $\ket{\psi}=U\ket{i_1,\cdots,i_n}$ and $\psi_1$ is the reduced density matrix of $\ket{0}\langle 0|$ on the first qubit.
\begin{theorem}
Constant depth quantum computation has no computational advantage over classical computation in decision problems.
\end{theorem}
We verify the statement by observing
\begin{align*}
\mathrm{tr}(\psi_1 \ket{0}\langle 0|)=\mathrm{tr}(\psi \ket{0}\langle 0|\otimes I_{2,\cdots n})=\langle 0^{\otimes n}| U^{\dag}(\ket{0}\langle 0|\otimes I_{2,\cdots n})U\ket{0^{\otimes n}}.
\end{align*}
For constant depth $U$, $U^{\dag}(\ket{0}\langle 0|\otimes I_{2,\cdots n})U$ applies non-trivially on at most a constant number of qubits, denoted as $B_{s}$, then
\begin{align*}
\mathrm{tr}(\psi_1 \ket{0}\langle 0|)=\langle 0\cdots 0| B_s\ket{0\cdots 0}
\end{align*}
This shows that one can compute $\mathrm{tr}(\psi_1 \ket{0}\langle 0|)$ efficiently, and thus confirm our claim for circuits.
The statement is also true for constant depth quantum operations rather than circuits.

\bigskip\noindent{\it Discussion:} Leveraging a single copy of an unknown quantum state for property learning represents a potent tool for analyzing NISQ quantum algorithms. This capability enables us to discard the assumption of possessing independent identical copies of the state.

The learnability with a single sample for low-complexity quantum states also serves as a technique for discriminating between such states. Immediately, it provides a strong lower bound on the trace distance between quantum states with low complexity.

We anticipate that the absence of advantage in decision problems and the ability to learn from a single sample for low-depth circuits can play a pivotal role in verifying and validating NISQ devices. As an example, one can employ the probabilistic method and the efficient computation of $\mathrm{tr}(\psi_1 \ket{0}\langle 0|)$ can provide a reasonable estimation of $O:=\frac{1}{n}Z_i$ through classical computation with high probability. These techniques can be a benchmark for the experimental realization of the low-depth computation and single copy estimation of $\mathrm{tr}(O\rho)$.

\bibliographystyle{apsrev4-1_with_title}
\bibliography{main}



\end{document}